*Working Paper*

# Distrust of social media influencers in America

**Shawn Berry, DBA[1]***

**June 3, 2024**

[1]William Howard Taft University, Lakewood CO, USA

*Correspondence: shawn.berry.8826@taftu.edu

**Abstract**  The popularity of social media influencers (SMI) as a means for businesses and causes to engage with the public and develop followers is undeniable. However, the use of SMI have been scrutinized due to various scandals that reflect poorly on brands and firms. Consequently, the distrust of SMI can create the potential for damage to a brand if audiences are not receptive to communications and messaging. This study (n=351) shares insights and findings of the apparent distrust of SMI that were discovered during the data analysis phase of my dissertation (Berry, 2024a). The study examines levels of trust and distrust toward SMI in the United States according to various demographic characteristics of respondents, specifically, age, gender, income level, education level, and region of the United States. Chi square analysis of the variables and a predictive model of trust of SMI are presented. Finally, recommendations and suggestions for future research are discussed.

## 1. Introduction

Social media influencers (SMI) are individuals that primarily act as spokespersons for a company, brand or cause to increase engagement with consumers through the creation of hype and awareness of a particular product or service offering through posts and video clips on social media platforms, and leveraging their own celebrity fan base and reputation. More precisely, Enke and Borchers (2021) "define social media influencers as third-party actors who have established a significant number of relevant relationships with a specific quality to and influence on organizational stakeholders through content production, content distribution, interaction, and personal appearance on the social web" (p.7). Despite the popularity of SMI, the usage of influencers remains a subject of criticism by many due to scandals and egregious behavior of influencers that subsequently reflect poorly on the brands and firms that contract their endorsements in detrimental way to the extent of degrading consumer trust and repelling purchase decisions for a brand (Reinikainen et al., 2021, von Mettenheim and Weidmann, 2023). As a result, the use of SMI have been called into question (Droz-dit-Busset, 2022; Hahl, 2023; Silva, 2020), and has been the subject of research in the topic of distrust of social media influencers which continues to evolve.

The availability of empirical findings with respect to the trust of SMIs to various socioeconomic variables in an American context are generally sparse. In their investigation of distrust of influencers, Zhang et al. (2024) observed that "participants exhibited varying patterns of trust and distrust across demographic characteristics" (p.1). With respect to the trust of SMI, Aikan and Ulas (2023) observed that although gender was not a statistically significant variable, income and education were statistically. Kadadha, Ermeley, and Amir (2022) observed "that males and females differ significantly in their rating of the impact of the brand, traditional celebrity, and social media influencers (SMI) on consumer's behavior" (p.328), and "that all the different household income categories differ significantly in their ratings to the impact of social media influencers (SMI) on the consumer's behavior" (p.331). While there does not appear to be empirical studies on the geographic variation of trust of SMI between regions of the United States, a study by Han and Balabanis (2023) suggest that there are variations in trust on a global basis with respect to Western versus Eastern cultures. Hautala (2019) observed that "age has contribution to causes of distrust towards social media influencers" (p.49), also noting that "age had also contribution to previous negative attitude towards social media influencers" (Hautala, 2019, p.49).

This study will share recent insights and findings of the apparent distrust of SMI held by respondents that were revealed during data analysis of my dissertation (Berry, 2024a). In particular, trust levels of SMI will be examined according to various demographic characteristics of respondents, namely, income, age, gender, level of education, level of income, and their region of residence in the United States. These findings are of particular value to marketers and researchers given the current trend for brands and firms to engage SMI as part of their marketing plans and strategies because it generally describes the overall sentiment of respondents to SMI. On a broad level, the findings should be seen as generalizable to the population at large, given the sample size. The study will look at the degree of association between the trust of SMI and the demographic characteristics of respondents. Finally, the study will offer guidance with respect to the use of SMI, and directions for future research.

## 2. Materials and Methods

This study is based on observations that were collected for my dissertation using an online questionnaire that was deployed on Amazon Mechanical Turk (mTurk) to male and female users residing in the United States (Berry, 2024a). Respondents were surveyed about how they used online reviews, along with additional attitudinal and belief data, and demographic information for classification purposes. Although the original desired sample was n=398, as reported in Berry (2024b), "11.8% of the 398 respondents were excluded due to not consenting to the study (3), self-reporting as not residing in the United States (8), attention check question failure (24), and identification of attempts to take the survey more than once (12)" (p.3). Thus, the usable sample was n=351 (Berry, 2024b).

The trust of SMI was extracted from responses to the trust of people construct that was employed in Berry (2024a). The trust of people instrument used a 5-point Likert scale wherein respondents were asked to rate the extent to which they trusted various kinds of individuals in society that are sources of influence (Berry, 2024b). Each of the 13 individual types ranged from those that were in authority roles, celebrities, those who are very familiar in everyday life, and those that are completely unknown to respondents (Berry, 2024b). As reported in Berry (2024b), "using Cronbach's alpha, the reliability of the instrument was 0.85 (Berry, 2024), which is considered to be high (Taber, 2018), and therefore, reliable" (p.3).

The data was coded for analysis according to the scheme as illustrated in Table 1 below.

**Table** 1

*Variable coding scheme*

| Variable name | Variable description | Variable type | Coding |
|---|---|---|---|
| Age | Age group of respondent, years of age | Categorical | 18-24 = 1<br>25-34 = 2<br>35-44 = 3<br>45-54 = 4<br>55 and over = 5 |
| Gender | Gender of respondent | Categorical | Female = 0<br>Male = 1<br>Non-binary = 2 |
| Income | Annual income level of respondent, USD | Categorical | Less than $30,000 = 1<br>$30,000-$49,999 = 2<br>$50,000-$69,999 = 3<br>$70,000 and over = 4 |
| Education | Education level of respondent | Categorical | Did not finish high school = 0<br>High school graduate = 1<br>Some college = 2<br>Bachelor's degree = 3<br>Master's degree = 4<br>Post-graduate or higher = 5 |
| Region | United States region of residence of respondent | Categorical | Middle Atlantic = 1<br>New England = 2<br>South Atlantic = 3<br>East South Central = 4<br>West South Central = 5<br>Mountain = 6<br>Pacific = 7 |
| Likert scores | Degrees of agreement or importance of behavioral factors to measure trust of people | Ordinal | Definitely distrust = 1<br>Somewhat distrust = 2<br>Neither trust nor distrust = 3<br>Somewhat trust = 4<br>Definitely trust = 5 |

**Source:** Berry (2024b), Table 1.

## 3. Results

Table 2 illustrates the trust of SMI according to the region of residence of respondents in the United States. The region with the greatest distrust of SMI was the South Atlantic region (22 definitely distrust and 28 somewhat distrust). The region with the least distrust of SMI was in New England. The greatest trust of SMI was in the Middle Atlantic region (2 definitely trust and 20 somewhat trust). While the greatest indifference of SMI was in the South Atlantic region, the least indifference toward SMI was in the New England and Mountain regions. The results of the chi square test of independence showed that there is no significant relationship between the trust of SMI and the region of residence of respondents, $X2$ (16, $N = 351$) $= 28.877$, $p = .225$.

**Table 2**
*Trust of social media influencers by region of United States*

| Region | 1 Definitely distrust | 2 Somewhat distrust | 3 Neither trust nor distrust | 4 Somewhat trust | 5 Definitely trust |
|---|---|---|---|---|---|
| Middle Atlantic (NY/NJ/PA) | 11 | 19 | 15 | 20 | 2 |
| New England (CT/ME/MA/NH/RI//VT) | 4 | 4 | 7 | 1 | 0 |
| South Atlantic (DE/DC/FL/GA/MD/NC/SC/VA/WV) | 22 | 28 | 27 | 10 | 3 |
| East South Central (AL/KY/MS/TN) | 12 | 17 | 18 | 10 | 0 |
| West South Central (AR/LA/OK/TX) | 17 | 11 | 12 | 7 | 3 |
| Mountain (AZ/CO/ID/MT/NV/NM/UT/WY) | 7 | 9 | 7 | 3 | 2 |
| Pacific (AK/CA/HI/OR/WA) | 14 | 15 | 9 | 5 | 0 |

**Source:** Data analysis, Berry (2024a).

Table 3 illustrates the trust of SMI according to age category of respondents. The age category with the greatest distrust of SMI were those respondents aged 25 to 34 (34 definitely distrust and 45 somewhat distrust). The age category with the least distrust of SMI were those respondents aged 55 and older (3 definitely distrust and 1 somewhat distrust). The greatest trust of SMI were those respondents aged 25 to 34 (3 definitely trust and 22 somewhat trust). The age category with the greatest indifference toward SMI were those respondents aged 25 to 34. The least indifference toward SMI was among those respondents aged 18-24 and those aged 55 and over. The results of the chi square test of independence showed that there is no significant relationship between the trust of SMI and the age category of respondents, $X2$ (16, $N = 351$) = 14.075, $p = .593$.

**Table 3**
*Trust of social media influencers by age category*

| Age category | 1 Definitely distrust | 2 Somewhat distrust | 3 Neither trust nor distrust | 4 Somewhat trust | 5 Definitely trust |
|---|---|---|---|---|---|
| 18–24 | 10 | 8 | 6 | 1 | 1 |
| 25–34 | 34 | 45 | 35 | 22 | 3 |
| 35–44 | 28 | 35 | 31 | 16 | 4 |
| 45–54 | 12 | 14 | 17 | 14 | 2 |
| 55 and older | 3 | 1 | 6 | 3 | 0 |

**Source:** Data analysis, Berry (2024a).

Table 4 illustrates the trust of SMI according to the level of income of respondents. The income categories with the greatest distrust of SMI was among those respondents earning less than $30,000 per year (24 definitely distrust and 30 somewhat distrust) and between $30,000 and $49,000 per year (29 definitely distrust and 24 somewhat distrust). The income category with the least distrust of SMI was among those respondents earning $50,000 and $69,999 per year (17 definitely distrust and 15 somewhat distrust). The greatest trust of SMI were those respondents earning between $30,000 and $49,000 per year (2 definitely trust and 26 somewhat trust). The income category with the greatest indifference toward SMI was among those respondents earning less than $30,000 per year. The least indifference toward SMI was among those respondents earning $50,000 and $69,999 per year. The results of the chi square test of independence showed that there is no significant relationship between the trust of SMI and the level of income of respondents, $X2$ (12, $N = 351$) = 17.691, $p = .125$.

**Table 4**
*Trust of social media influencers by income level*

| Age category | 1 Definitely distrust | 2 Somewhat distrust | 3 Neither trust nor distrust | 4 Somewhat trust | 5 Definitely trust |
|---|---|---|---|---|---|
| Less than $30,000 | 24 | 30 | 42 | 16 | 5 |
| $30,000–$49,999 | 29 | 24 | 25 | 26 | 2 |
| $50,000–$69,999 | 17 | 15 | 11 | 5 | 1 |
| $70,000 and over | 17 | 25 | 17 | 9 | 2 |

**Source:** Data analysis, Berry (2024a).

Table 5 illustrates the trust of SMI according to the level of education of respondents. The education categories with the greatest distrust of SMI was among those respondents with some college education (38 definitely distrust and 40 somewhat distrust) and those possessing a bachelor's degree (32 definitely distrust and 35 somewhat distrust). The education category with the least distrust of SMI was among those respondents that did not finish high school (0 definitely distrust and 1 somewhat distrust) and those with a postgraduate or higher education (3 definitely distrust and 4 somewhat distrust). The greatest trust of SMI were those respondents holding master's degrees (3 definitely trust and 19 somewhat trust). The income category with the greatest indifference toward SMI was among those respondents with some college education. The least indifference toward SMI was among those respondents with a postgraduate or higher education and among those respondents that did not finish high school (0). The results of the chi square test of independence showed that there is a significant relationship between the trust of SMI and the level of education of respondents, $X2$ (20, $N$ = 351) = 42.176, $p$ = .003.

**Table 5**

*Trust of social media influencers by level of education*

| Level of education | 1 Definitely distrust | 2 Somewhat distrust | 3 Neither trust nor distrust | 4 Somewhat trust | 5 Definitely trust |
|---|---|---|---|---|---|
| Did not finish high school | 0 | 1 | 0 | 1 | 0 |
| High school graduate | 11 | 13 | 16 | 6 | 0 |
| Some college | 38 | 40 | 32 | 15 | 3 |
| Bachelor's degree | 32 | 35 | 26 | 15 | 4 |
| Master's degree | 3 | 10 | 14 | 19 | 3 |
| Postgraduate or higher | 3 | 4 | 7 | 0 | 0 |

**Source:** Data analysis, Berry (2024a).

Table 6 illustrates the trust of SMI according to the gender of respondents. The greatest distrust of SMI was among female respondents (52 definitely distrust and 75 somewhat distrust). The least distrust of SMI was among non-binary respondents (2 definitely distrust and 1 somewhat distrust). The greatest trust of SMI was among female respondents (5 definitely trust and 35 somewhat trust). The greatest indifference toward SMI was among female respondents. The least indifference toward SMI was among male respondents. The results of the chi square test of independence showed that there is a significant relationship between the trust of SMI and the level of education of respondents, $X2$ (8, $N = 351$) = 12.487, $p = .131$.
.

**Table 6**
*Trust of social media influencers by gender*

| Gender | 1 Definitely distrust | 2 Somewhat distrust | 3 Neither trust nor distrust | 4 Somewhat trust | 5 Definitely trust |
|---|---|---|---|---|---|
| Female | 52 | 75 | 73 | 35 | 5 |
| Male | 33 | 27 | 22 | 21 | 5 |
| Non-binary | 2 | 1 | 0 | 0 | 0 |

**Source:** Data analysis, Berry (2024a).

The data was analyzed using linear regression to model the trust of SMI according to the demographic characteristics. The results appear in Table 7. Although most of the chi square tests resulted in most variables not being significantly associated with the trust of SMI, the regression analysis shows that gender was the only demographic variable that was not statistically significant. Trust levels of SMI appear to decrease with the level of income. Trust levels of SMI increase with the level of education and age. Given the sign and coefficient on the region variable, the regional effect of trust of SMI does not appear to be as strong as that for level of income but significant, nonetheless. The gender variable implies that men and binary individuals appear to have distrust of SMI. In general, the size of the coefficients of the age and education level variables appear to be the biggest factors that contribute to the increase in trust of SMI.

**Table 7**

*Regression model of respondents levels of trust of SMI (dependent variable: smitrust)*

|           | Estimate | Std. Error | t value | Pr(|t|) | Significance |
|-----------|----------|------------|---------|---------|--------------|
| Intercept | 2.238    | 0.244      | 9.183   | <0.001  | (***)        |
| Gender    | -0.069   | 0.119      | 2.445   | 0.561   |              |
| Income    | -0.172   | 0.056      | -3.100  | 0.002   | (**)         |
| Region    | -0.081   | 0.030      | -2.677  | 0.008   | (**)         |
| Education | 0.187    | 0.060      | 3.114   | 0.002   | (**)         |
| Age       | 0.151    | 0.062      | 2.445   | 0.015   | (*)          |

$^*p < .05$. $^{**}p < .01$. $^{***}p < .001$.

**Source:** Data analysis, Berry (2024a)

**4. Discussion**

The findings broadly suggest that regardless of demographic characteristics, the majority of respondents hold levels of distrust of SMI, whether somewhat or definitely distrusting them. Although there are interesting regional variations of trust in the United States, there is no statistically significant association between region of residence and levels of trust of SMI. However, the regional variation in distrust and indifference should still generally imply that, within the United States, consumer trust of SMI is not the same for every region. The level of education of respondents was found to be statistically significant in chi square analysis and predictive modeling, confirming the observation of Aikan and Ulas (2023). The findings suggest that as the level of education increases, some groups of those with some college and undergraduate education to be distrustful possibly because of skepticism or some other resistance to trust, and those with graduate education appear more trusting. Contrary to Aikan and Ulas (2023) and Kadadha, Ermeley, and Amir (2022), gender was not found to be statistically significant with respect to trust levels of SMI in both chi square and predictive modeling analysis. However, level of income was found to be a statistically significant predictor of trust of SMI, confirming the observations of Aikan and Ulas (2023) and Kadadha, Ermeley, and Amir (2022). Females were found to be most distrustful of SMI. Although the age category of respondents was not found to be statistically significant with respect to trust levels of SMI in chi square analysis, it was found to be a statistically significant predictor of trust of SMI. Thus, the observations of Hautala (2019) that age is a significant variable is confirmed, however, not confirming the negative influence of age on trust. In general, respondents between 24 and 44 years of age comprised the largest contingent that were distrustful of SMI, notably those between 24 and 34 years of age.

As a result, these findings have implications for business practice. First, with respect to the findings for various age groups, more careful consideration must be given to the appropriate communications approach to reach a given demographic group, and probably should not consider SMI as a default or go-to approach. Given that no one group of people or region of the United States demonstrates a great amount of trust for SMI, these facts should be interpreted as a cue to better understand the communication preferences and receptivity of consumers in a more granular and nuanced way. Therefore, messaging must be keyed to each group in order to be believed and trusted. Second, the findings imply that using SMI as a generic communication approach would be a waste of money since people do not hold high levels of trust, including high levels of indifference toward SMI. Therefore, if SMI are being seriously considered as part of a marketing plan, this author recommends that their effectiveness should be tested with groups and panels before implementing in order to mitigate costs, and to address any confirmed distrust and non-belief of SMI by test subjects, ensuring to get meaningful feedback about the reasons for distrust. Finally, the reasons for wanting to use SMI in the first place must be honestly explored, including how they align to a corporate vision, and their alignment to the values of the target audiences, if any. Furthermore, if a product or brand is facing challenges due to a lack of consumer demand or waning brand image, this author suggests that marketers honestly study if SMI would fix what may be an underlying issue, such as quality, poor internal organization that yields poor customer service, or other internal problems that cannot be eliminated with a SMI, as examples. .

## 5. Directions for future research

First, future research efforts should attempt to understand the trust drivers of people that neither trust nor distrust SMI as this kind of indifference may imply that messaging and communications will not be received. Second, efforts should be made to determine the granularity of trust. Given the findings of this study that broadly suggest that various demographic groups are distrustful of SMI, this suggests that other factors are at play which must be understood with respect to the activation of trust of SMI or any person that is tasked with communicating a brand message to ensure message reception.

## 6. Limitations

Every study is not perfect, and therefore, a reflection upon the limitations of a study must be conducted. First, as acknowledged in Berry (2024b), the use of Amazon Mechanical Turk, while efficient and expeditious, is not without its flaws or challenges. Therefore, the author acknowledges the drawbacks of convenience sampling using mTurk, even with the use of best practices (Berry, 2024b). Second, as the topic of my dissertation was about the effects of Google restaurant reviews written by AI on consumers (Berry, 2024a), the topic of this study was inspired by the new insights that were discovered from the analysis of my trust constructs and instruments. Thus, since the responses to the trust instruments by respondents represent the extent to which they trust various individuals in different roles, it is assumed that respondents generally understood the meaning of trust in this context. The author acknowledges that, on this basis, a formal definition of trust was not offered to respondents in the questionnaire, assuming that the meaning of the word should not be ambiguous. Finally, although the author acknowledges that this study is not exhaustive, the findings are offered as important new insights that were synthesized from my dissertation that incrementally add to the body of knowledge.

## 7. Patents

There are no patents resulting from the work reported in this manuscript.

## 8. Funding

This research received no external funding.

## 9. Conflicts of Interest

The authors declare no conflict of interest.

## 10. Declaration of generative AI in scientific writing

The author declares that generative AI tools were not used in the writing or research of this article.

Shawn Berry, DBA                                                                                                                                    13 of 13